\documentclass[letterpaper,10pt,nofootinbib,aps,tightenlines,prl,twocolumn]{revtex4}
\usepackage{amsmath,amsfonts,amssymb}
\usepackage{mathrsfs} 
\usepackage{graphicx}
\usepackage{color}
\usepackage[english]{babel}
\usepackage[utf8]{inputenc}

\begin{document}

\title{Using a Primordial Gravitational Wave Background to Illuminate New Physics}
\author{Robert R. Caldwell${}^1$}
\author{Tristan L. Smith${}^2$}
\author{Devin G. E. Walker${}^1$}
\affiliation{${}^1$Department of Physics \& Astronomy, Dartmouth College, 6127 Wilder Laboratory, Hanover, NH 03755 USA \\
${}^2$Department of Physics \& Astronomy, Swarthmore College, Swarthmore, PA 19081 USA}
\date{\today}

\begin{abstract}

A primordial spectrum of gravitational waves serves as a backlight to the relativistic degrees of freedom of the cosmological fluid. Any change in the particle physics content, due to a change of phase or freeze-out of a species, will leave a characteristic imprint on an otherwise featureless primordial spectrum of gravitational waves and indicate its early-Universe provenance. We show that a gravitational wave detector such as the Laser Interferometer Space Antenna would be sensitive to physics near $100$~TeV in the presence of a sufficiently strong primordial spectrum. Such a detection could complement searches at newly proposed 100km circumference accelerators such as the Future Circular Collider at CERN and the Super Proton-Proton Collider in China, thereby providing insight into a host of beyond Standard Model issues, including the hierarchy problem, dark matter, and baryogenesis.
 
\end{abstract}

\maketitle

Changes in the relativistic degrees of freedom of the cosmological fluid of the early Universe leave an imprint on a primordial spectrum of superhorizon and subhorizon gravitational waves (GWs).  The physical mechanism is easy to understand: a speedup in the expansion rate, as when the fluid cools past the rest mass of any species, will slightly dilute all subhorizon gravitational radiation relative to the background \cite{Bennett:1985qt}; superhorizon waves are frozen, however, and are unaffected by any hiccups in the rate of expansion. The final, processed spectrum shows a series of steps downward, going from low to high frequency, corresponding to changes in the relativistic degrees of freedom \cite{Watanabe:2006qe}. Our goal is to quantify the size of the steps in the GW spectrum, and show that a new path to physics beyond the Standard Model may be within reach of the Laser Interferometer Space Antenna (LISA) \cite{Audley:2017drz}.

We require the existence of a primordial stochastic GW background (SGWB) at a detectable amplitude in order to access new physics beyond the Standard Model. Yet this may not be so outrageous, for several reasons. First, recent theoretical work has identified a wide class of early-Universe scenarios in which a strongly amplified, blue-tilted GW spectrum is produced \cite{Cook:2011hg,Maleknejad:2011jw,Adshead:2013qp,Namba:2015gja,Maleknejad:2016qjz,Dimastrogiovanni:2016fuu,Adshead:2016omu,Caldwell:2017chz,Fujita:2018ehq,Maleknejad:2018nxz}. Hence, the existence of a SGWB to serve as a backlight is within the realm of current thinking about the early Universe. Second, following on the success of the LISA Pathfinder mission \cite{Armano:2016bkm,Armano:2018kix}, LISA has recently rebooted and a design analysis is in progress. This means a mHz-band GW experiment that is sensitive enough to place meaningful bounds on a SGWB may become a reality in the early 2030s \cite{Audley:2017drz}. The frequencies probed by LISA would correspond to changes in the relativistic degrees of freedom of the cosmological fluid at temperatures spanning $T \sim 10^3$ - $10^7$~GeV. This range of energies includes the reach of the high energy Large Hadron Collider (HE-LHC) as well as a proposed 100km circumference Future Circular Collider at CERN (FCC-hh) or the Super Proton-Proton Collider (SppC) in China that would achieve energies up to 100 TeV \cite{Benedikt:2018ofy,Tang:2015qga}. Hence, synergy between LISA and future accelerators could provide insight into the hierarchy problem, dark matter, supersymmetry or composite theories, but also completely new territory. There is good reason to suspect new physics beyond the Standard Model lurks at these energies \cite{Arkani-Hamed:2015vfh,Mangano:2017tke}. And whereas particle physics experiments are sensitive only to new physics that couples to the Standard Model, this ``backlight effect" is sensitive to all gravitating degrees of freedom, light and dark. 

Previous work that investigated the degree to which a space laser interferometer can determine the thermal history of the early Universe focused on gathering information about the equation of state of the early Universe \cite{Seto:2003kc, Boyle:2005se,Watanabe:2006qe,Saikawa:2018rcs} or the post-inflationary reheat temperature \cite{Nakayama:2008ip,Kuroyanagi:2011fy,Kuroyanagi:2014nba}. There is much ongoing work considering early Universe phase transitions, either for the GWs they themselves produce in the case of a strongly first-order transition \cite{Caprini:2015zlo,Kuroyanagi:2018csn}, or the effect that a weaker, crossover transition may have on an inflationary spectrum \cite{Jinno:2011sw}. Our work is distinct in that we consider the ability of LISA to distinguish the modulation of a primordial spectrum due to rather conservative speculations of new TeV-scale physics.

Discovery of a primordial stochastic background would be profound. Upon detecting an irreducible noise, however, one cannot immediately tell if it is an astrophysical foreground from unresolved sources, or a primordial relic. It is expected that astrophysical modeling of GW sources can be translated into frequency and directional information, as a template to remove known foregrounds. But the identification of any residual background remains a challenge, particularly if the residual is an otherwise featureless power law. The phenomenon we investigate is a clear indicator of primordial provenance: a SGWB emitted across a range of times, particularly one of inflationary origin, should display the tell-tale steps in amplitude that mark it as a primordial spectrum.


{\it Gravitational Waves.}--- 
We consider a linearized description of weak GWs $h_{ij}$ propagating in an expanding spacetime $ds^2 = a^2(\tau)(-d\tau^2 + (\delta_{ij} + h_{ij})dx^i dx^j)$. The equation of motion for the Fourier amplitude $\tilde h_{ij}(\tau,\vec k)$ is
\begin{equation}
\tilde h_{ij}'' + 2 \frac{a'}{a}\tilde h_{ij}' + k^2 \tilde h_{ij} = 16 \pi G a^2 \Pi_{ij}
\label{eqn:hdiff}
\end{equation}
where $'$ indicates derivative with respect to conformal time and $\Pi_{ij}$ is the anisotropic shear tensor. The comoving frequency $f$ is related to the wavenumber $k = 2 \pi f$.  Although the shear of the cosmic fluid gives rise to some important effects (e.g. Ref.~\cite{Watanabe:2006qe}), we will ignore its possible contribution for now, thereby setting the right hand to zero. We can further simplify the evolution by separating the frequency- and time-dependent amplitude $h(k,\tau)$ from the polarization tensor: $\tilde h_{ij} = h_P e^P_{ij}$ for $P=+,\,\times$. Hereafter we drop the polarization index for simplicity. The comoving expansion rate $a'/a$ distinguishes two regimes of behavior. For superhorizon modes, $k \ll a'/a$, the dominant solution for $h$ is a constant.  For subhorizon modes, $k \gg a'/a$, the solution is oscillatory. The transfer function relating the initial amplitude $h_i$ at early times to the present-day amplitude, as a function of scale, depends sensitively on the details of the intervening expansion history. In a radiation-dominated background, with $a \propto \tau$, the analytic solution is $h= h_i [\sin k(\tau-\tau_i)+ k \tau_i \cos k(\tau-\tau_i)]/(k \tau)$ where we assume initial conditions that are consistent with inflation, $h=h_i$, $h'=0$ at some suitably early time such that $k \tau_i \ll 1$. The energy density in GWs is $\rho_{GW} =  \langle h'_{ij}(\tau,\vec x) h'^{ij}(\tau,\vec x)\rangle /{32 \pi Ga^2}$ where the angle brackets indicate averaging over a time interval much greater than the period of oscillation. Inserting the above analytic solution for $h$ into the expression for energy density, we obtain the spectral density, $\Omega_{GW} \equiv {d(\rho_{GW}/\rho_c)}/{d\ln f}$, where $\rho_c$ is the present-day critical density. For cosmic evolution that departs from radiation domination, however, we numerically solve Eq.~(\ref{eqn:hdiff}) subject to the same initial conditions to find the effect on the spectral density.


{\it Cosmic Fluid.}---
The description of the radiation-dominated epoch is based on the free-field thermodynamics of a collection of noninteracting bosons and fermions in thermal equilibrium \cite{Kolb:1990vq}:
\begin{eqnarray}
\rho &=& \sum_j \frac{g_j}{2 \pi^2} \int_{m_j}^\infty dE \frac{E^2 \sqrt{E^2-m_j^2}}{e^{E/T_j}-s_j} \label{eqn:rho}\\
p &=& \sum_j \frac{g_j}{6 \pi^2} \int_{m_j}^\infty dE \frac{(E^2-m_j^2)^{3/2}}{e^{E/T_j}-s_j}. \label{eqn:prs}
\end{eqnarray}
The sum is over all particle species of mass $m_j$, $g_j$ is the multiplicity or degrees of freedom, and $s_j = \pm 1$ for bosons/fermions. Our notation allows the temperatures for different species to differ, but in equilibrium we expect all temperatures to be the same. At high temperatures, above the rest mass energy of all species $T \gg m_j$, the energy density and pressure are $\rho = 3p= g_* \pi^2 T^4/30$, and $g_*$ is the effective number degrees of freedom in the relativistic gas.  

\begin{figure}[t]
\begin{center}
\includegraphics[width=0.85\linewidth]{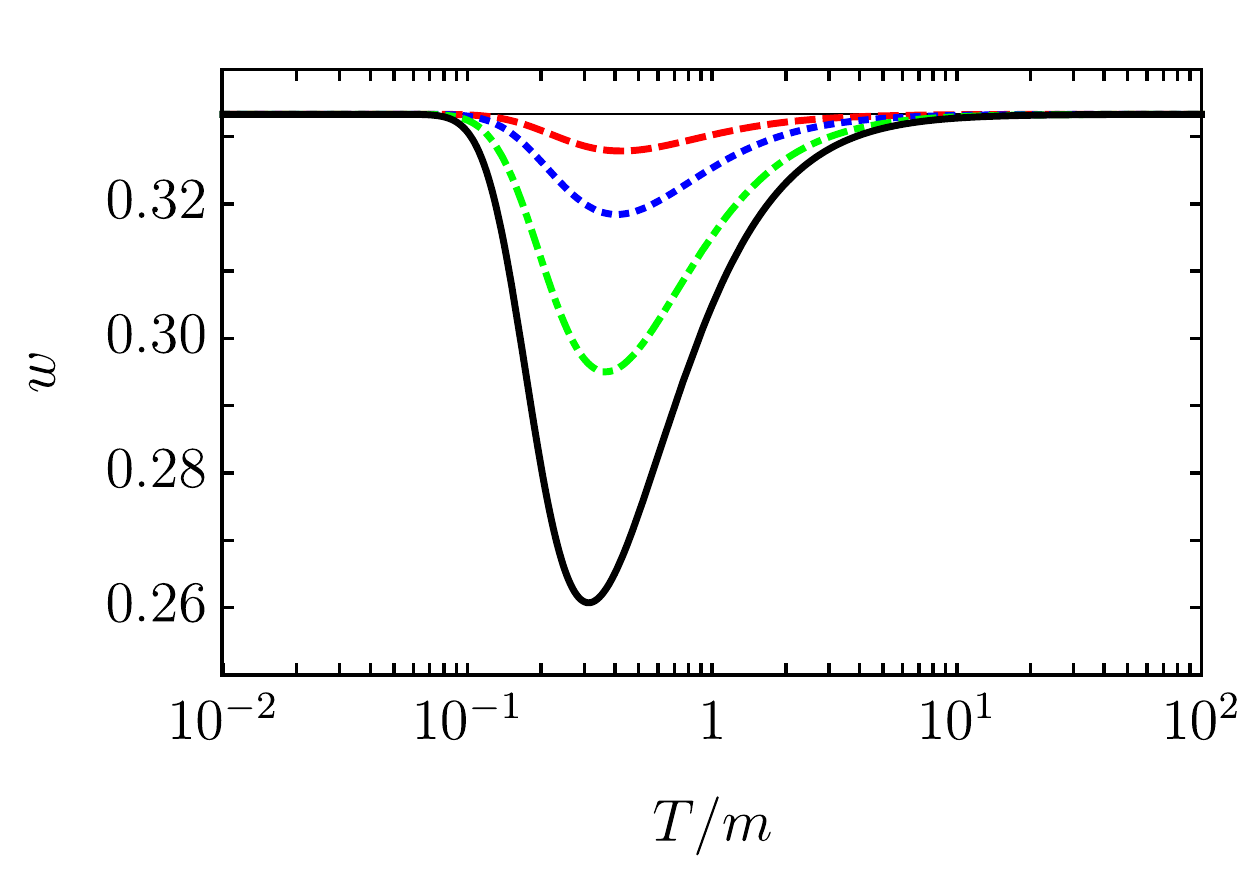}
\includegraphics[width=0.85\linewidth]{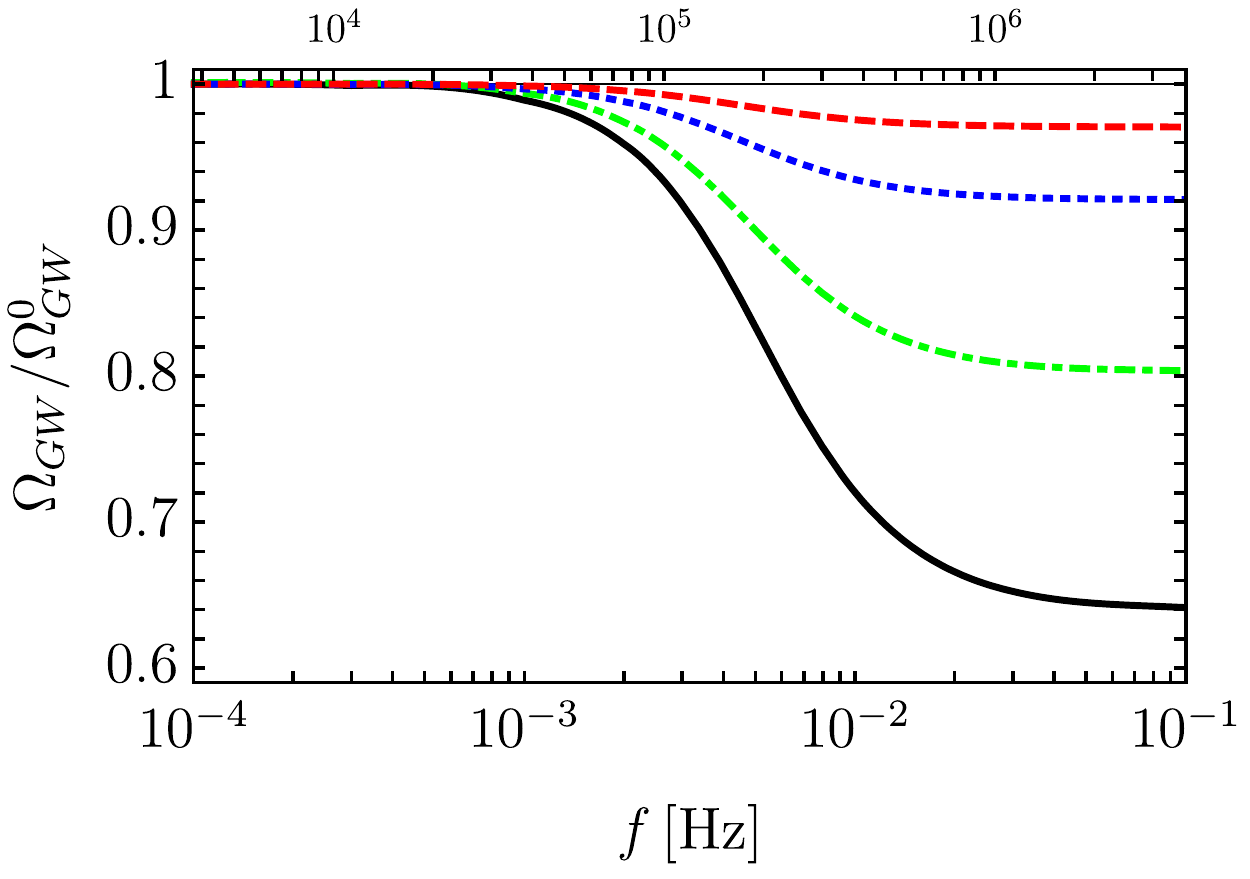}
\end{center}
\caption{(Top) The equation of state of the cosmic fluid, consisting of $106.75$ relativistic degrees of freedom as in the Standard Model, and $g_*= 0,\, 10,\, 30,\, 100,\, 300$ degrees of freedom, descending, as a function of $T/m$. (Bottom) The step feature in a SGWB for $g = 10,\, 30,\, 100\,\ 300$ degrees of freedom of mass $m = 4 \times 10^5$~GeV. The Standard Model with $g_{SM}= 106.75$ degrees of freedom has been assumed. For lower or higher masses, the curves shift left or right, respectively. The top axis label gives the translation to the temperature of the cosmic fluid in GeV, according to which $2 \pi f = H a/a_0$.}
\label{fig:w}
\end{figure}

Now consider an individual species in thermal equilibrium with the rest of the fluid. As the temperature drops below the mass, the pressure given by Eq.~(\ref{eqn:prs}) drops slightly more rapidly than the energy density. As the particle species thereby becomes non-relativistic, the equation of state of the cosmic fluid temporarily drops below the relativistic case $p/\rho = 1/3$. This is also indicated by a positive trace of the stress-energy tensor $\Theta = \rho - 3 p$, which displays a spike relative to $T^4$. (See Fig.~2 of Ref.~\cite{Caldwell:2013mox}.) The slight disturbance in the equation of state affects the redshift rate of the cosmic fluid and the Hubble damping in the GW equation. This is the origin of the effect we consider. 
 
To model the impact of the thermal history on the spectrum of GWs, we evolve Eq.~(\ref{eqn:hdiff}) in the background of a cosmic fluid with $g_{SM}=106.75$ relativistic degrees of freedom, plus $g$ additional degrees of freedom at a collective mass $m$. Eqs.~(\ref{eqn:rho}-\ref{eqn:prs}) are used to build the background cosmology. Examples of the equation of state history as functions of temperature are shown in the top panel of Fig.~\ref{fig:w}. We calculate the spectral density for a sequence of modes spanning present-day frequencies $f \in [10^{-4},\, 10^{-1}]$~Hz. Upon studying many cases in which $g$ and $m$ are varied, for $g \in [0,10^3]$ and $m \in [10^3,\,10^7]$~GeV, we find the resulting feature in the spectrum is well fit by the function $\Omega_{GW}(f) = \Omega_{GW}^0(f) F(f;\, g,\, m)$ where
\begin{equation}
F(f;\, g,\, m) = \frac{1 - \epsilon(g) \tanh[\ln f/f_0(m) ]}{1 + \epsilon(g)}.
\label{eqn:F}
\end{equation}
Here, $\epsilon= (1-\Delta)/(1+\Delta)$ where $\Delta \simeq (1 + g/g_{SM})^{-1/3}$ and $2 \pi f_0 = H a/a_0|_{T\simeq m/b}$ with $b = 2.2/\Delta$ determined empirically. Illustrated in the right panel of Fig.~\ref{fig:w} are examples of the resulting step-like feature or break in the spectral density, which we seek to detect. We find that a mass in the vicinity of $100~$TeV corresponds to a feature at mHz frequencies.

This simple parametrization also provides an effective description for a crossover transition, as occurs for the electroweak Higgs symmetry breaking transition as well as for QCD at the confinement transition. In both cases, the effect on the expansion rate is well described using free-field thermodynamics. In the case of the electroweak transition, the mass and degrees of freedom of participating species are known, so that the effect on the cosmic expansion may be calculated. For the QCD transition, lattice simulations are required to determine the critical temperature and strength of the conformal anomaly, which can be translated into a mass $m$ and effective degrees of freedom $g$. Beyond the Standard Model, we expect that the phenomenological impact of a crossover in an SU(N) can also be described using Eq.~(\ref{eqn:F}), where $g$ scales as the appropriate power of the number of charges of the gauge field and the coupled fermion families. Hence, a crossover transition in the vicinity of $100~$TeV will also leave an imprint at mHz frequencies. 

The effect of an out-of-equilibrium decay of a non-relativistic species can also be accommodated within our model. Consider a species X with mass $m_X$ that drops out of equilibrium and freezes out at an abundance $Y_X$. Following the blueprint for thermal dark matter, this non-relativistic species will eventually dominate over the radiation. However, if it subsequently decays at a rate $\Gamma_X$ into Standard Model radiation which thermalizes with $g_X$ degrees of freedom, this species can drive a departure from pure radiation-domination and produce the same step-like feature in a SGWB. In this case, we can still use Eq.~(\ref{eqn:F}), but now $\Delta = 1- g_X/g_{SM}$ and $2 \pi f_0 = H a/a_0|_{\Gamma_X = H}$. The abundance is related as $Y_X \simeq \tfrac{3}{4}m_X^{-1}(\Delta^{-1}-1)[90\Gamma_X^2 M_P^2/\pi^2 g_{SM}]^{1/4}$. In this case, a decay rate $\Gamma_X$ that is roughly $(100~{\rm TeV})^2/M_P$ would leave a mHz imprint. Since the particle species would be non-relativistic after dropping out of equilibrium, $m_X$ must be $10^4$~TeV or larger.


{\it LISA.}--- 
The Laser Interferometer Space Antenna is a proposed mission by the European Space Agency (ESA) to detect long wavelength GWs. LISA is three Michelson interferometers, consisting of a trio of spacecraft in an equilateral triangle configuration; each spacecraft, carrying a pair of isolated test masses, laser and optics bench, is in a freely falling, Earth-trailing orbit around the Sun.  The distance between spacecraft is $L=2.5 \times 10^6$~km, which sets the characteristic frequency in the mHz range. The mission requirements prescribe a sensitivity range spanning the interval $[0.1,\, 100]$~mHz \cite{Audley:2017drz}.

\begin{figure}[b]
\begin{center}
\includegraphics[width=0.95\linewidth]{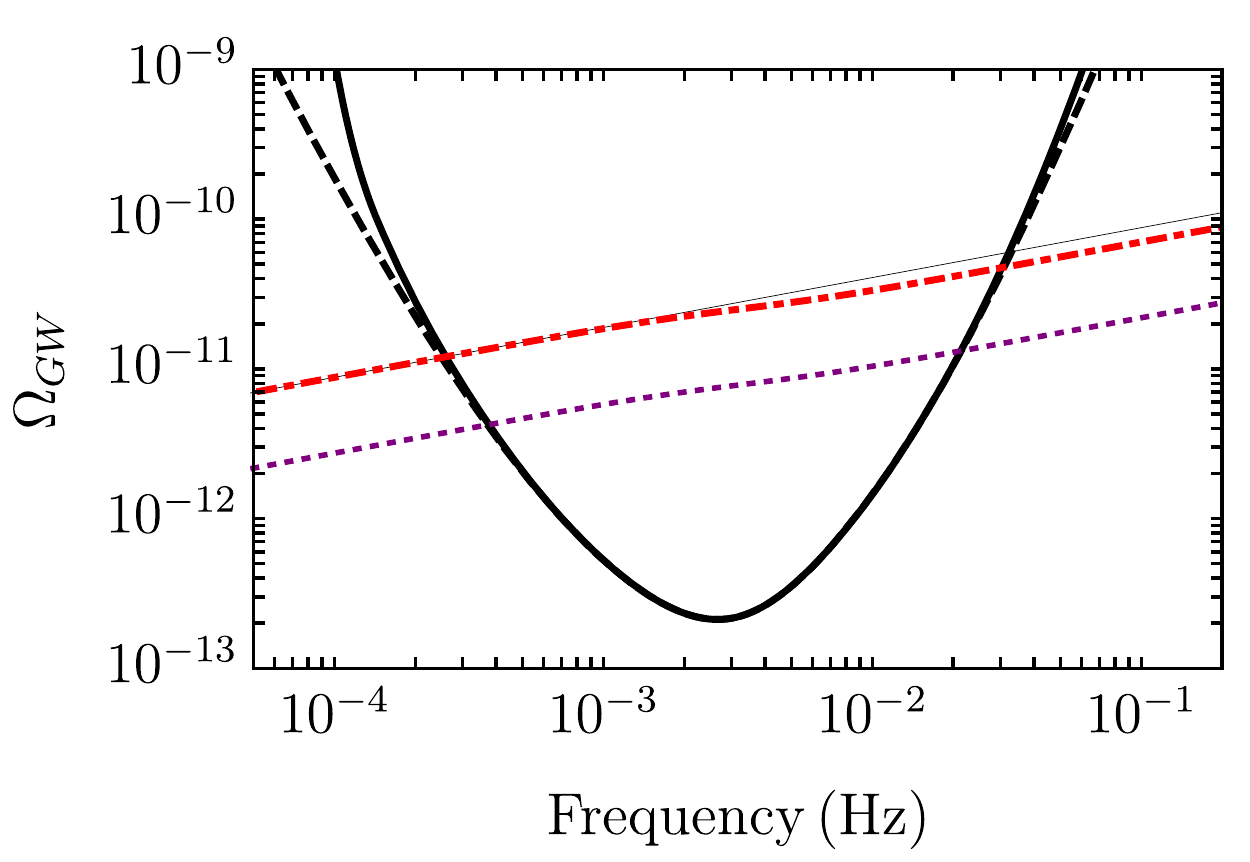}
\caption{Tilted SGWBs consistent with current bounds that feature a step ($g=100$, $m=100$~TeV) are shown on a sensitivity curve for LISA (black solid). A SGWB without the step (thin black line) is included for contrast. For the upper, red dashed case, SNR=9 and the Fisher forecast measurement uncertainty is $\sigma_g=15,\, \sigma_m=24$~TeV; the feature in the lower, purple dotted case is at the threshold of detectability, with SNR=3 and $\sigma_g=50,\, \sigma_m=75$~TeV.}
\label{fig:LISAsens}
\end{center}
\end{figure}

The sensitivity of LISA to a SGWB may be estimated by considering the signal to noise ratio of the optimal statistic 
\begin{equation}
    {\rm SNR}^2 = \sum_{a={\rm A,E}} T \int_{f_{\rm min}}^{f_{\rm max}} df \left( \frac{{\cal S}_a(f)}{{\cal N}_a(f)}\right)^2.
    \label{eqn:snrformula}
\end{equation}
where ${\cal S}$ is the signal covariance matrix, ${\cal N}$ is the noise power spectrum, dominated by acceleration and optical metrology shot noise, and $T$ is the observation time \cite{Romano:2016dpx}. The sum is over the two independent autocorrelation modes, labelled $a={\rm A,\, E}$. We implicitly assume that a third mode ${\rm T}$ is used to characterize and clean the noise from the ${\rm A}$, ${\rm E}$-modes \cite{Hogan:2001jn,Adams:2010vc}. The signal due to a SGWB is
\begin{equation}
{\cal S}_{a}[\Omega_{GW}(f)] = \mathcal{R}_{a}(f) \, |W(f)|^2 \, I[\Omega_{GW}(f)]
\end{equation}
where $\mathcal{R}_{a}$ is the response of the detector geometry to an isotropic distribution of GWs, $W$ is a factor that accounts for the time-delay interferometry (TDI) used to mitigate the effects of laser power noise and satellite drift, and the intensity is $I[x(f)] = {3 H_0^2 x(f)}/{4 \pi^2 f^3}$ \cite{Maggiore:1999vm}. These expressions are identical for both ${\rm AA}$ and ${\rm EE}$ modes. We set the threshold for detection of a SGWB to an integrated signal-to-noise ratio SNR=3 for three years observational data. The resulting sensitivity curve for LISA to a featureless, scale-free spectrum is shown in Fig.~\ref{fig:LISAsens}. Any power-law SGWB that crosses above the sensitivity curve is, in principle, detectable \cite{Thrane:2013oya}. In this simplistic analysis, we assume that astrophysical foregrounds from unresolved galactic sources may be distinguished for their anisotropic distribution and cleanly removed \cite{Adams:2013qma,Robson:2017mhm}.

To determine the sensitivity to the step in the spectrum, we adapt a matched filter approach and consider a $\chi^2$-inspired SNR, replacing ${\cal S}_a$ in Eq.~(\ref{eqn:snrformula}) by ${\cal S}_a[\Omega_{GW}(f)] - {\cal S}_a[\Omega^N_{GW}(f)]$. This closely resembles the statistic developed in Ref.~\cite{Kuroyanagi:2018csn}. In the preceding expression, $\Omega_{GW}(f)$ is the SGWB in the presence of $g$ additional degrees of freedom of mass $m$. We model this as $\Omega_{GW}(f) = \Omega_{GW}^0(f) F(f;g,m)$, where $\Omega_{GW}^0(f) = A_{GW} (f/f_*)^{n_T}$ and $F$ is given by Eq.~(\ref{eqn:F}). The other term, $\Omega^N_{GW}(f) \equiv \Omega_{GW}^{0\prime}(f)$, is the spectrum without the feature; the prime indicates that we allow different values of $A_{GW}$ and $n_T$ in the reference spectrum, which we marginalize over. We use this statistic to determine whether the difference between the spectra with and without the feature is large enough, relative to the noise, to be detectable. We minimize the SNR with respect to $\Omega^N_{GW}$ to find the value that best fits the spectrum with the step. If $g$ is too small, or if $m$ is too extreme for the feature to lie within the LISA band, then we expect $\Omega_{GW}(f)$ to be indistinguishable from a featureless spectrum. We set a modest threshold ${\rm SNR}>3$ for detectability of the step in the spectrum. 
 

We can also use a Fisher analysis to determine how well a GW observatory can measure a step in the SGWB spectrum \cite{Tegmark:1996bz,Kuroyanagi:2018csn}. The covariance in the $a={\rm A,E}$ interferometer signals is
\begin{equation}
{\bf C} = \frac{1}{2}\left[\mathcal{S}_a(f) + \mathcal{N}_a(f)\right] \delta_{ab}.
\end{equation}
Assuming that the data is drawn from a Gaussian distribution, the Fisher information matrix is given by \cite{Tegmark:1996bz}
\begin{eqnarray}
F_{\alpha \beta} &=& \frac{1}{2}{\rm Tr} \left[{\bf C}^{-1} \frac{\partial{\bf C}}{\partial \theta_\alpha}{\bf C}^{-1} \frac{\partial{\bf C}}{\partial \theta_\beta}\right],\\
&\simeq& \frac{1}{2}T \sum_{a=A,E}\int_{f_{\rm min}}^{f_{\rm max}} \frac{\frac{\partial\mathcal{S}_a(f)}{\partial \theta_\alpha}\frac{\partial\mathcal{S}_a(f)}{\partial \theta_\beta}}{\left[\mathcal{N}_a(f)+\mathcal{S}_a(f)\right]^{2}}df,
\end{eqnarray}
where $\theta_\alpha$ are the parameters used to model the SGWB, and again we have assumed that the instrumental noise can be completely characterized by monitoring the Sagnac ($T$-mode) signal. The inverse of the Fisher matrix is the parameter covariance matrix giving us estimates for their uncertainties (see, e.g., Ref.~\cite{2009arXiv0906.4123C}).   

As before we model the SGWB as a power-law with a step so that the spectrum can be described by four parameters: $A_{GW}$, $n_T$, $g$, $m$; we take the pivot frequency $f_* \equiv 1/(2\pi L)$.  In order to ensure that all of the elements of the Fisher matrix are of similar order (so that it is well-conditioned) we parameterize the SGWB amplitude by $\ln A_{GW}$ and $\ln m$ where $m$ is in units of $10^5$ GeV. We find that both the SNR and Fisher approaches produce the same estimated uncertainties in the model parameters. 

\begin{figure}[t] 
\includegraphics[width=0.95\linewidth]{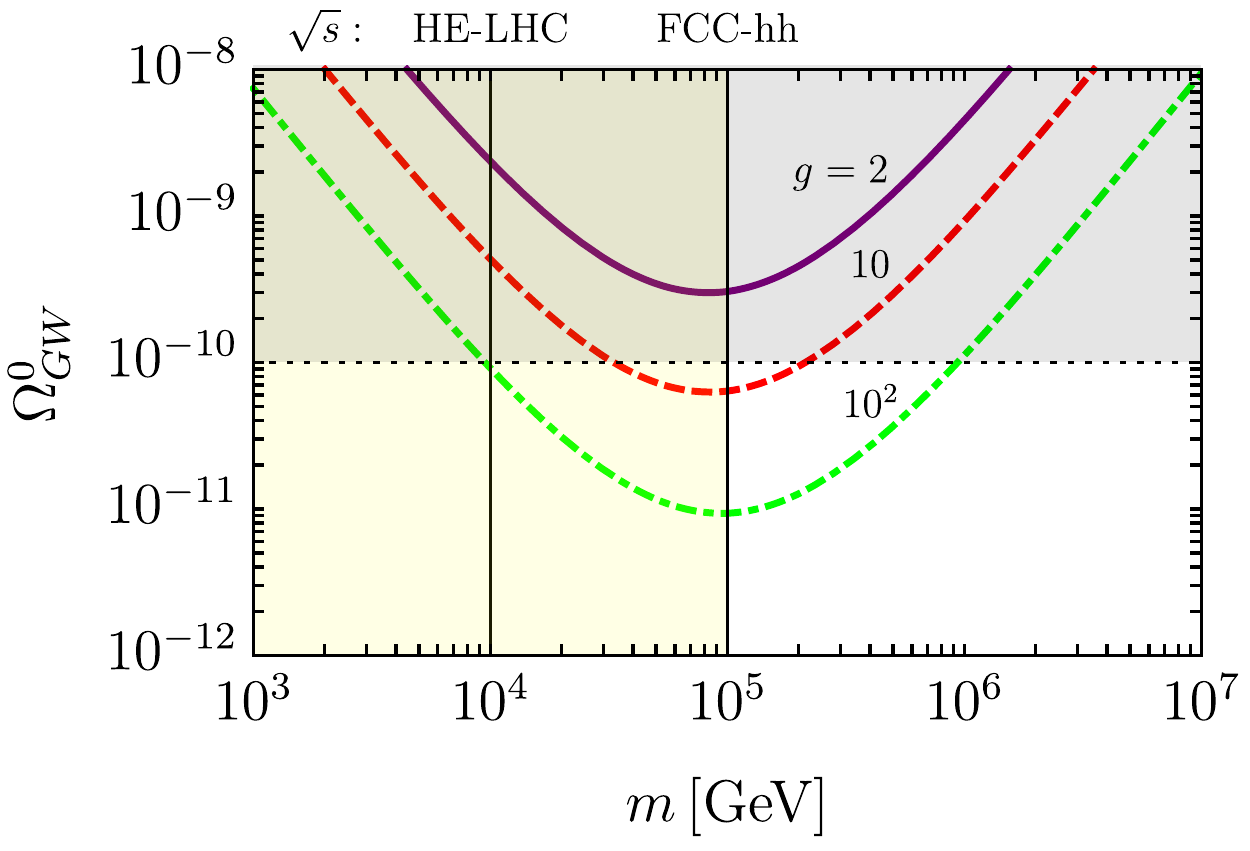}
\caption{The threshold $\Omega_{GW}^0$ needed to identify the backlight effect for fixed degrees of freedom is shown as a function of the mass. The yellow shaded region shows the range of energies to be explored by proposed accelerators. The grey shaded region shows the level of SGWB excluded by current observations (in the absence of new physics at higher energies).}
\label{fig:constraints}
\end{figure}

The sensitivity of LISA to the thermal history of the Universe is summarized in Fig.~\ref{fig:constraints}. First, we see that the threshold $\Omega_{GW}$ to identify the backlight effect decreases monotonically as the number of degrees of freedom increases. However, the relative gain in sensitivity diminishes as the floor of the LISA sensitivity window is reached. Second, LISA is most sensitive to effects that correspond to frequencies near a few mHz, which translates into a shift in degrees of freedom at a mass scale near $100$~TeV.

 
{\it The inflationary SGWB.}--- 
In a universe filled with matter and radiation the GW background at LISA frequencies predicted by inflation is given by \cite{Boyle:2005se}
\begin{equation}
\Omega^0_{GW} (f) = \frac{rA_s}{24} \Omega_r \left(\frac{f}{f_{\rm cmb}}\right)^{n_T}.
\end{equation}
We evaluate the spectral density as follows. Using the temperature and polarization measurements of the 2018 \texttt{Planck} data release \cite{Aghanim:2018eyx} as well as data from the Keck Array and BICEP2 collaborations \cite{Array:2015xqh} the scalar perturbation amplitude is $10^9A_s=2.100 \pm 0.030$, the tensor to scalar ratio is constrained to be $r < 0.07$ (95\% C.L.) so we define $r_7 \equiv r/0.07$, and the CMB pivot frequency is $f_{\rm cmb} = 1.94 \times 10^{-17}\ {\rm Hz}$; the radiation energy density consisting of photons at a temperature of $T_{\rm cmb} = 2.7$ K and three nearly massless neutrinos is $\Omega_r h^2 = 4.15 \times 10^{-5}$; the Hubble constant is measured to be approximately $h\simeq 0.7$ \cite{Aghanim:2018eyx,2018ApJ...855..136R}. If the primordial SGWB is scale invariant (i.e., $n_T=0$) then in the absence of any particle physics effects, the amplitude at mHz frequencies is 
\begin{equation}
\Omega^0_{GW}(f) \leqslant 5  \times 10^{-16}r_7,
\end{equation}
which is well out of reach of LISA. However, the situation is different if the spectrum is strongly blue tilted, as has been proposed recently (see, e.g., Refs.~\cite{Cook:2011hg,Adshead:2013qp,Namba:2015gja,Dimastrogiovanni:2016fuu,Adshead:2016omu,Caldwell:2017chz,Fujita:2018ehq}). In this case, a primordial signal may be within reach. We can use a variety of upper limits on $\Omega_{GW}$ coming from measurements of the CMB, pulsar timing arrays, LIGO, and indirect constraints from the contribution of the short-wavelength SGWB to the radiative energy density of the universe \cite{Smith:2006nka} to arrive at a bound $n_T<0.39 - 0.04\log_{10}(r/0.07)$ at the 95\% confidence level \cite{Lasky:2015lej}. Using these constraints, the upper limit to the SGWB in the mHz range is given by 
\begin{equation}
\Omega^0_{GW}(f) \leqslant 1.8 \times 10^{-10} r_7^{0.4} \left(\frac{f}{3~{\rm mHz}}\right)^{0.39 - 0.04\log_{10}r_7}.
\end{equation}
Comparing with Fig.~\ref{fig:LISAsens}, the idealized, peak LISA sensitivity to a stochastic background is several orders of magnitude better than the current upper limit, leaving ample room for discovery.


{\it Beyond the Standard Model (BSM) Scenarios.}---  
The backlight effect can be used to probe the new physics at multi-TeV to PeV temperatures predicted in a variety of BSM scenarios. For example, symmetry-breaking phase transitions are a staple of model building which, if detected, would indicate new fundamental laws of matter. Likewise, out of equilibrium decays are a generic feature in a variety of BSM scenarios, including ones that explain the hierarchy problem, baryogenesis, and also dark matter. A simple example is a dark photon generated by adding a $U(1)'$, spontaneously broken gauge symmetry to the Standard Model. In this scenario the dark photon, $Z'$, is coupled to the hypercharge gauge boson via kinetic mixing,
$L_\mathrm{mix} = \frac{\epsilon}{2}\,F'_{\mu\nu} B^{\mu\nu}$. Here $B$ and $F'$ are the field strength tensors for the photon and dark photon, respectively.  The $Z'$ has a mass of $m_{\gamma'} \gg m_Z$ with the coupling~\cite{Babu:1997st} 
\begin{equation}
    \mathcal{L}_\gamma = - \frac{e}{s_w c_w} \overline{\psi}_i \gamma^\mu \left(g_v^{(i)} + g_a^{(i)}\gamma_5 \right) \psi_i A'_\mu,
\end{equation}
where 
\begin{align}
    g_v^{(i)} &=  \frac{c_w s_w g'}{e \sqrt{1 - \epsilon^2} } \left( \frac{1}{2} T_3^i - s_w^2 Q^i\right) \\
    &+ \frac{\epsilon\,s_w}{\sqrt{1 - \epsilon^2}} \left( \frac{1}{2}T_3^i - Q^i \right) \nonumber \\
    g_a^{(i)} &= \frac{1}{2}T_3^i - \frac{\epsilon\,s_w}{2\sqrt{1 - \epsilon^2}} T^i_3. 
\end{align}
This coupling accounts for the fact that electroweak symmetry has not been broken at the time of interest. The subsequent decay of the dark photon into fermionic electroweak multiplets occurs with the width
\begin{eqnarray}
    \Gamma &=& \sum_{i\, = \,\mathrm{SM\,fermions}} \frac{N_c}{12\pi} 
    \frac{m_\gamma\,e^2}{s_w^2c_w^2}\left(g_v^{(i)\,2} + g_a^{(i)\,2} \right)\sqrt{1 - 4 y_i}\,\cr
    &\times& \left[1 + 2 y_i\left(\frac{g_v^{(i)\,2} - 2\, g_a^{(i)\,2}}{g_v^{(i)\,2} + g_a^{(i)\,2}} \right) \right]
\end{eqnarray}
where $y_i = m_i^2/m_{\gamma'}^2$, respectively.  For simplicity, we also assume the dark Higgs boson, which is responsible for spontaneously breaking the $U(1)'$, is heavier than the dark photon mass. The kinetic mixing parameter, which controls the decay lifetime, can be arbitrarily small. Consequently, a long-lived particle of mass $m_X \gtrsim 10^6$~GeV with decay constant $\Gamma_X \simeq 1.3 \times 10^{-8}$~GeV and abundance $Y_X \simeq 1.8\times 10^{-2} (m_X/10^6~{\rm GeV})^{-1}$ would have the appearance of a crossover transition with $100$ degrees of freedom at $120$~TeV, right in the center of the LISA window.  This translates to $\epsilon \lesssim 10^{-2}$ for the model above. 

If the sole addition to the Standard Model up to these mass scales is a single massive species, such as a thermal dark matter candidate at the limit of the unitarity bound $\lesssim 100$~TeV \cite{Griest:1989wd}, then detection would require a background $\Omega_{GW}^0 \gtrsim 5 \times 10^{-10}$. This threshold is inconsistent with the current bound; however, if there are further changes in the particle content at higher energies, then the bound may be softened. Continued improvement will restrict attention to blue-tilted backgrounds and limit the ability to resolve the number of species.


{\it Discussion.}---  
Detection of GWs across a wide range of frequencies can provide crucial information about the underlying source. We have shown that the mHz frequency band has the potential to open up a new window on beyond Standard Model physics that is complementary to searches at newly proposed 100km circumference accelerators such as the Future Circular Collider at CERN and the Super Proton-Proton Collider in China. There are further implications of the backlight effect across frequencies spanning the CMB ($f\sim 10^{-18}$~Hz) to LIGO/Virgo and future ground-based observatories \cite{Sathyaprakash:2012jk,Evans:2016mbw} ($f\sim 100-1000$ Hz). If a primordial power-law spectrum is detected in the CMB, then we can expect the amplitude at the low-frequency end of the LISA window to be suppressed relative to a naive power-law extrapolation. The combined effects of neutrino viscosity, the non-relativistic transition of electrons, and the QCD and EW Higgs crossover transition serve to lower the amplitude of higher-frequency waves by a factor of $\sim 0.4$. (See Fig.~5 of Ref.~\cite{Watanabe:2006qe}.) Similarly, due to unknown physics beyond $100$~TeV, the high frequency bound on a SGWB may be weakened when extrapolated back to the LISA window.

Other concepts for space-based GW detectors may probe even higher energies than LISA. The Big Bang Observer is a similar constellation of three spacecraft, with separation $L=5 \times 10^7$~m, therefore sensitive to higher frequencies or higher energy scales in the thermal history of the Universe. Under the assumption of position and acceleration noise sensitivity $10^{-17}~{\rm m}/\sqrt{\rm Hz}$ and $3 \times 10^{-17}~{\rm m}/{\rm s}^2 \sqrt{\rm Hz}$ \cite{Crowder:2005nr}, we find that the sensitivity curves have the same shape as for LISA, but shifted by two orders of magnitude higher in mass.

The results derived here may also apply to a GW spectrum emitted by a network of cosmic strings or other scaling sources \cite{Figueroa:2012kw}. A scaling network emits a scale free spectrum of GWs during the radiation era. Loops radiate at frequencies $f_n = n/(\alpha t)$ for $n=1,\,2,\,...$ until they evaporate away, where simulations suggest $\alpha$ is $10^{-3}$ or smaller. If the loop lifetime is sufficiently short, and the power in higher harmonics drops steeply, then to first approximation loops radiate into their fundamental mode for a duration that is much shorter than a Hubble time. If these conditions hold, then changes in the degrees of freedom of the cosmological fluid should be imprinted on the radiation spectrum of loops: radiation from all loops that are present when a particle species becomes non-relativistic will be slightly diluted; radiation yet to be emitted from loops that have not yet formed will not be diluted \cite{Bennett:1985qt,Caldwell:1991jj}. Since $\alpha \ll 1$, events at a temperature $T$ will impart features at the higher frequency $2 \pi f = \alpha^{-1} H a/a_0 |_T$. This means the thermal history of the Standard Model could lie in the LISA band. However, we caution that the step feature will be smeared due to both extended loop lifetime and power in higher harmonics \cite{Ringeval:2017eww,Blanco-Pillado:2017oxo,Cui:2017ufi,Cui:2018rwi}. Since there remains considerable uncertainty regarding loop lifetime and spectra \cite{Wachter:2016hgi,Hindmarsh:2017qff}, as well as the conditions under which the network forms \cite{Bettoni:2018pbl}, we leave investigation of the detectability of this effect for cosmic string spectra for future study. 

\acknowledgments

The work of RRC is supported in part by DOE grant DE-SC0010386. 

\bibliography{paper.bib}

\end{document}